\documentclass[epj,onecolumn]{svjour}
\usepackage{graphics,amsmath}
\usepackage[utf8]{inputenc}
\usepackage[english]{babel}
\begin{document}
\title{Comment on the paper "The third-order perturbed Korteweg-de Vries equation for shallow water waves with a non-flat bottom" by  M. Fokou, T.C. Kofan\'e, A. Mohamadou and E. Yomba,\\ Eur. Phys. J. Plus, 132, 410 (2017)}
\author{Piotr Rozmej\inst{1} \and Anna Karczewska\inst{2}
}                     
%
%
\institute{Faculty of Physics and Astronomy,
University of Zielona G\'ora, Szafrana 4a, 65-246 Zielona G\'ora, Poland \and Faculty of Mathematics, Computer Science and Econometrics, University of Zielona G\'ora, Szafrana 4a, 65-246 Zielona G\'ora, Poland}
\date{Received: date / Revised version: date}
%
\abstract{
The authors of the paper  "The third-order perturbed Korteweg-de Vries equation for shallow water waves with a non-flat bottom" \cite{FKMY} claim that they have derived the full third order perturbed KdV equation for the  case of uneven bottom. We show that the authors' derivation is not consistent due to the fact that they took into account only some of the third order corrections but not all of them. Moreover, we show that a consistent third order perturbed Korteweg-de Vries equation for shallow water waves with a non-flat bottom cannot be derived for a general form of bottom function.
} 
\maketitle
\section{The extended KdV equation for uneven bottom}

It is widely known that the ubiquitous Korteweg - de Vries equation, in the case of shallow water waves, is obtained by a perturbation approach which is first order in small parameters and assumes a flat bottom for the fluid container.
The second order equation, also for this even bottom case, was derived first by Marchant and Smyth in 1990 \cite{MS90} and named the \emph{extended KdV}. 

In the papers \cite{KRR,KRI} we have derived, for the first time, the nonlinear equation describing shallow water gravity waves for uneven bottom  
\begin{align} \label{kdv2B} 0 &=  \eta_t+\eta_x + \frac{3}{2}\alpha \eta\eta_x +\frac{1}{6}\beta \eta_{3x} - \frac{3}{8}\alpha^2\eta^2\eta_x  \hspace{5ex}   \\ &  +\alpha\beta\left(\frac{23}{24}\eta_x\eta_{2x}+\frac{5}{12}\eta\eta_{3x}\right) 
+\beta^2\frac{19}{360}\eta_{5x} \nonumber \\ & +\beta\delta \frac{1}{4}\!
\left(\!h_{3x}\eta+ h_{2x}\eta_x-h_{x}\eta_{2x}+h\eta_{3x}-\frac{2}{\beta}\left(h_{x}\eta+h\eta_{x}\right)\!\right).\nonumber
\end{align}
The equation (\ref{kdv2B}) has been obtained 
by a perturbation method up to the second order in small parameters $\alpha,\beta,\delta$.

The authors of the discussed paper cite our equation as \cite[Eq.~(2)]{FKMY}, which in their paper does not have the correct form. In \cite[Eq.~(2)]{FKMY} the coefficient of the linear term $\eta_{3x}$ is $\frac{1}{3}\beta$ instead of the correct $\frac{1}{6}\beta$. Moreover, all coefficients in the bracket $\beta\delta(\cdots)$ slightly differ from the correct ones. The authors write these terms as
\begin{equation} \label{bede}
\beta\delta\!\left(\!\frac{13}{72}h_{3x}\eta\! +\! \frac{1}{8}h_{2x}\eta_x \!-\!\frac{7}{24}h_{x}\eta_{2x} \!-\!\frac{17}{72}h\eta_{3x}\! -\!\frac{2}{3\beta}\left(h_{x}\eta \!+\!h\eta_{x}\right)\!\!\right)\!,
\end{equation}
whereas the correct form is 
\begin{equation} \label{bedeC}
\beta\delta\!\left(\!\frac{1}{4}\left(h_{3x}\eta \!+\! h_{2x}\eta_x\right) \!-\!\frac{1}{4}\left(h_{x}\eta_{2x} \!+\!h\eta_{3x}\right) \!-\!\frac{1}{2\beta}\left(h_{x}\eta \!+\!h\eta_{x}\right)\!\right).
\end{equation}
[Compare with  \cite[Eq.~(35)]{KRR} or \cite[Eq.~(18)]{KRI}.]
 
In our opinion most parts of the paper \cite{FKMY} are incorrect since the derivation presented there is {\bf inconsistent}, which we show below.

The inconsistency has arisen since the authors have taken the boundary condition at the variable bottom in the form \cite[Eq.~(14)]{KRI}, that is,
\begin{equation} \label{botBC}
 \phi_z=\beta\delta (h_x\phi_x) \quad \mbox{for} \quad z=\delta h(x).
\end{equation}
This is correct up to the second order in small parameters and it was enough in our papers  \cite{KRR,KRI} for consistent derivation of second order wave equation for a non-flat bottom.  The equation (\ref{botBC}) together with the Laplace equation \linebreak \cite[Eq.~(10)]{FKMY} allows us to express all odd order functions  
$\phi^{(m)}$ through $\phi^{(0)}$, $h$ and their derivatives according to
\begin{equation} \label{botBC1}
 \phi^{(1)}=\beta\delta \left(h_x\phi_x^{(0)}+h\phi_{2x}^{(0)}\right) 
\end{equation}
and
\begin{equation} \label{botBC2}
\phi^{(2k+1)} =\frac{(-\beta)^{k+1}}{(2k+1)!}\phi^{(1)}_{2kx}.
\end{equation}

A consistent perturbation approach of the third order requires, however, to take into account the bottom boundary condition to the same third order. As we pointed out in \cite[Eq.~(14)]{KRR} this condition takes the form of a complicated differential equation
\begin{equation} \label{botBC3}
 \phi^{(1)}-\beta\delta  \left(h_x\phi_x^{(0)}+h\phi_{2x}^{(0)}\right) -\beta\delta^2 \left(h\,h_x\phi_x^{(1)}+\frac{1}{2}h^2\phi_{2x}^{(1)}\right) =0 
\end{equation}
which does not supply a simple expression for $\phi^{(1)}$ such as~(\ref{botBC1}).
For arbitrary bottom function $h(x)$ it is not possible to find an explicit form of $\phi^{(1)}$ as a function of $\phi^{(0)}, h$ and their derivatives. Therefore consistent derivation of a KdV-type equation, third order in all three small parameters $\alpha,\beta,\delta$ is not possible. This was the reason why in papers \cite{KRR,KRI} we limited our study to the second order perturbation approach.

The authors of the critiqued paper \cite{FKMY} are not aware of the latter fact. They take bottom boundary condition (\ref{botBC1}) for granted and insert it into the equation for velocity potential \cite[Eq.~(15)]{FKMY}. Subsequently they proceed as recommended by Burde and Sergyeyev \cite{BS13}, taking into account terms up to third order in small parameters. This procedure is in some part third order and in another part second order and therefore {\bf totally inconsistent}.

Moreover, the authors repeat several technical errors. The coefficient in front of the linear term $\eta_{3x}$ is correctly written as $\frac{1}{6}\beta$ in the KdV equation \cite[Eq.~(1)]{FKMY} (when surface tension is neglected), but written wrongly as $\frac{1}{3}\beta$ in equations \cite[Eq.~(2)]{FKMY} and \cite[Eq.~(40)]{FKMY} and then again correctly in the equation \cite[Eq.~(42)]{FKMY} of the paper. The terms of the order $\beta\delta$ shown here in (\ref{bede}) appear with an incorrect coefficient not only in \cite[Eq.~(2)]{FKMY} but also in both \cite[Eq.~(40)]{FKMY} and \cite[Eq.~(42)]{FKMY}.

The most astonishing error consists in the term $\frac{1}{3}\eta_{3x}$ in equations \cite[Eq.~(2)]{FKMY} and \cite[Eq.~(40)]{FKMY}. This term is well known in KdV, where it appears correctly as $\frac{1}{6}\eta_{3x}$. None of the perturbative approaches of higher order can change it. It is to be questioned how this term was obtained in the discussed paper. We can not explain it, since in one of the previous papers \cite{FKMY16} the same team of authors present the third order KdV equation where this term is correct. This term is correct in \cite[Eq.~(42)]{FKMY}, as well.

\section{Comparison of some numerical results}

All these inconsistencies and mistakes make the numerical results presented in \cite{FKMY} questionable. On the other hand the authors show their numerical results for rather small values of parameters $\alpha,\beta$ 
and for relatively short times of evolution. Hence the higher order effects can be small and do not have enough time to show up, particularly when $\delta$ is small, too. 
In order to compare the authors' numerical results with the evolution according to second order equation, given in our papers \cite{KRR,KRI}, we decided to recalculate some of the presented cases with our own code.

\subsection{Case of ascendant bottom}

We focus on the simplest cases presented in Figs.~2 and 3 of the paper \cite{FKMY}. In \cite[Fig.~2]{FKMY} the authors show the time evolution of the initial KdV soliton, given by \cite[Eq.~(43)]{FKMY} with $\tau=0$. Time evolution is calculated numerically according to \cite[Eq.~(40)]{FKMY} with parameters $\alpha=\beta=0.1$ and $\delta=0.5$. The bottom function is taken as $h_+(x)=\pm\frac{1}{2}(\tanh(0.055(x-55))+1)$.

\begin{figure}[hbt]
\begin{center}
\resizebox{0.99\columnwidth}{!}{\includegraphics{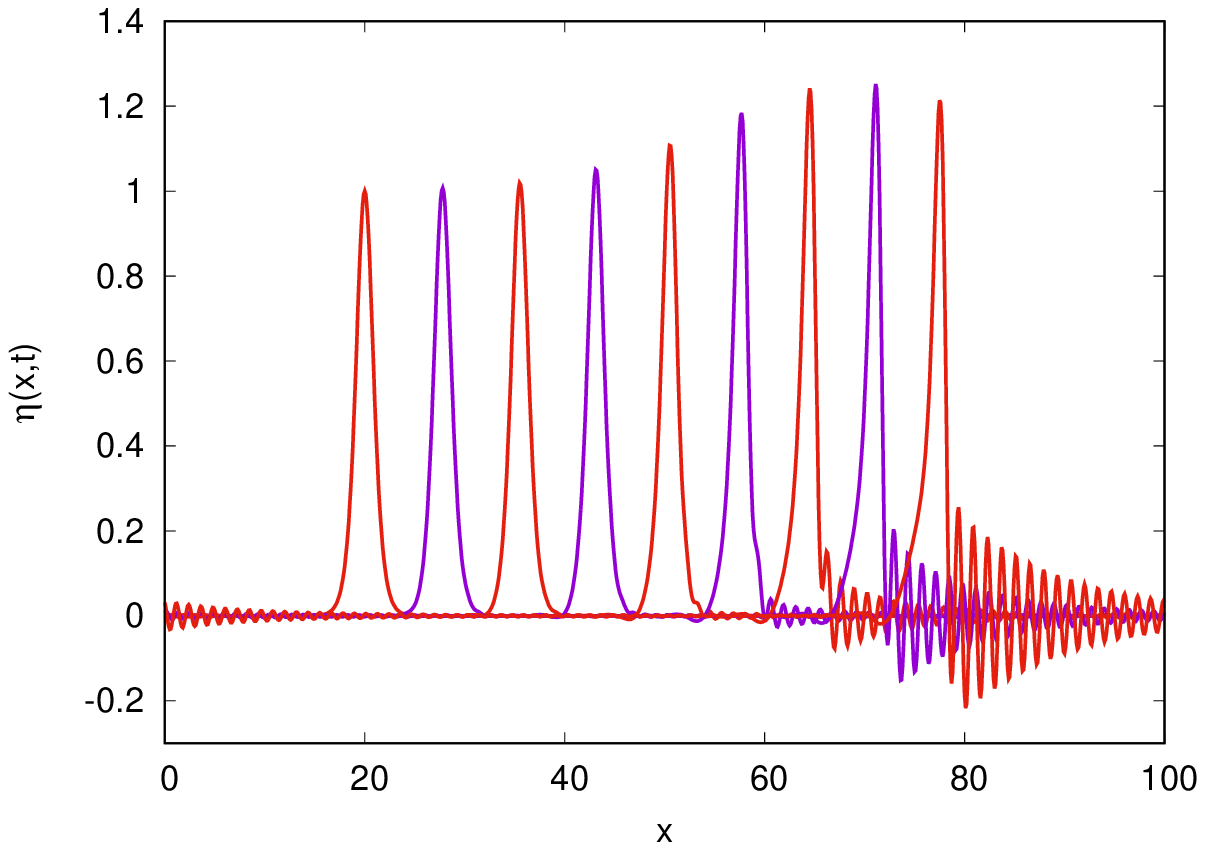}}
\end{center}
\caption{Time evolution of the initial KdV soliton obtained with second order equation derived in \cite{KRR,KRI}. The bottom function $h_+(x)$ and the values of parameters $\alpha,\beta,\delta $ are the same as in the Fig.~2 of \cite{FKMY}.} \label{ab1d5+}
\end{figure}

In Fig.~\ref{ab1d5+} we display the time evolution of the same initial KdV soliton, the same values of $\alpha,\beta,\delta$, the same interval $x\in [0,100]$ and the same space and time steps $\Delta x, \Delta t$ as in \cite[Fig.~2]{FKMY} but obtained according to a second order equation derived by us in \cite{KRR,KRI}. Only time instances corresponding to particular profiles may be slightly different since this information is not supplied in the paper \cite{FKMY}. However, since solitons cover similar distances the comparison of both numerical time evolutions is possible.

In \cite[Fig.~2]{FKMY} the amplitude of the wave increases from 1 to $\approx 1.8$ and profiles are distorted (by secondary wave) behind the main part only. 
In our calculations presented in Fig.~\ref{ab1d5+} we observe soliton radiation in front of the main wave and no distortions behind it. This radiation which occurs when a soliton enters a shallower region is  known in shallow water theory, see, e.g. \cite{Daw,Kuz,GrimPoF}.
In our case the amplitude increases only from 1 to $\approx 1.3$ at $x\approx 70$. The next decrease of the main wave amplitude is just the effect of the radiation mentioned above (which is relatively big since $\delta=0.5$ parameter, contrary to $\alpha=\beta=0.1$ is not small).

\begin{figure}[tbh]
\begin{center}
\resizebox{0.99\columnwidth}{!}{\includegraphics{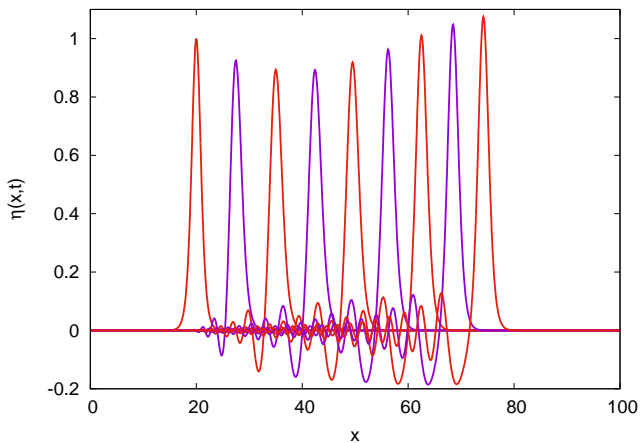}}
\end{center}
\caption{Time evolution of the initial KdV soliton obtained with evolution according to \cite[Eq.~(40)]{FKMY} but limited to second order terms, that is, equation (\ref{kdv2d}). The bottom function is $h_+(x)$.} \label{ab1d5+FKMY}
\end{figure}

\begin{figure}[tbh]
\begin{center}
\resizebox{0.99\columnwidth}{!}{\includegraphics{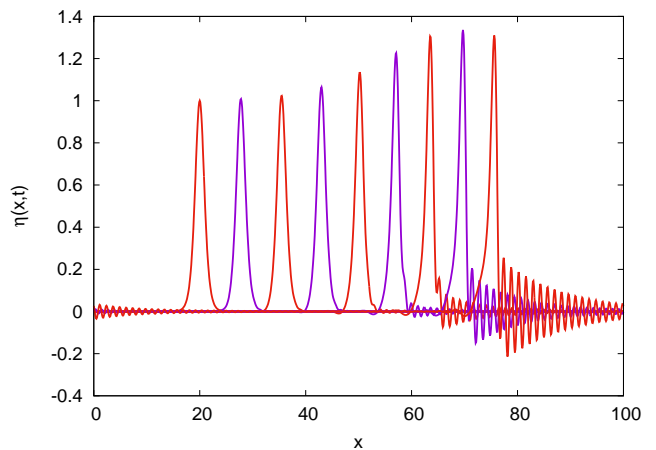}}
\end{center}
\caption{Time evolution of the initial KdV soliton obtained with evolution according to  equation (\ref{kdv2d}) in which coefficient $\frac{1}{3}\beta$ is replaced by the correct one $\frac{1}{6}\beta$. The bottom function is $h_+(x)$.} \label{ab1d5+FKMYcB}
\end{figure}

One wonders why the authors do not obtain the soliton radiation in  \cite[Fig.~2]{FKMY}.
In order to find the answer we made the following test. We inserted the coefficients of the equation \cite[Eq.~(40)]{FKMY} into our code and ran the time evolution according to the equation
\begin{align} \label{kdv2d}   \eta_t+\eta_x + \frac{3}{2}\alpha \eta\eta_x +\frac{1}{3}\beta \eta_{3x} \hspace{24ex} &   \\ - \frac{3}{8}\alpha^2\eta^2\eta_x   + \!\alpha\beta\!\left(\!\frac{23}{24}\eta_x\eta_{2x}\!\!+\!\!\frac{5}{12}\eta\eta_{3x}\! \right)\!+\!\beta^2\!\frac{19}{360}\eta_{5x} &\nonumber \\  +\beta\delta \!\left(\!\frac{13}{72}h_{3x}\eta\! +\! \frac{1}{8}h_{2x}\eta_x \!-\!\frac{7}{24}h_{x}\eta_{2x} \!-\!\frac{17}{72}h\eta_{3x} \right. &\nonumber  \\ - \left. \frac{2}{3\beta} h_{x}\eta -\frac{2}{3\beta} h\eta_{x}\!\right)\! &=0 \nonumber \end{align}
that is, the  equation \cite[Eq.~(40)]{FKMY} limited to second order.
The result of this numerical integration is displayed in Fig.~\ref{ab1d5+FKMY}.
This shows that soliton radiation preceding the main wave has disappeared. 
Is it the effect of the wrong coefficient in front of $\eta_{3x}$ or is it the effect of wrong coefficients in $\beta\delta$ terms? To answer this question we restored the coefficient $\frac{1}{3}\beta$ to $\frac{1}{6}\beta$ in (\ref{kdv2d}) and ran the code once more. The result is presented in Fig.~\ref{ab1d5+FKMYcB}.

It is clear that the presence of the correct $\frac{1}{6}\beta\eta_{3x}$ which is crucial already for KdV (that is, first order equation) restores known properties of the soliton profile when it approaches a shallowing. The equation \cite[Eq.~(40)]{FKMY}  gives so poor a time evolution since it is already {\bf wrong in the first order term}. The incorrect coefficients in $\beta\delta (\cdots)$ terms are only slightly different from the correct ones (see (\ref{bede}) and (\ref{bedeC})). Therefore they cause much smaller deviations from correct time evolution given in Fig.~\ref{ab1d5+} and the results shown in Fig.~\ref{ab1d5+FKMYcB} are almost the same as those in Fig.~\ref{ab1d5+} .
It is seen, however, that the  incorrect coefficients in $\beta\delta (\cdots)$ terms together with not fully consistent third order terms induce a much faster increase of the solitons amplitude in \cite[Fig.~2]{FKMY} than that observed by us in the evolution according to the second order equation.

\subsection{Case of sloping bottom}

\begin{figure}[htb] 
\begin{center}
\resizebox{0.99\columnwidth}{!}{\includegraphics{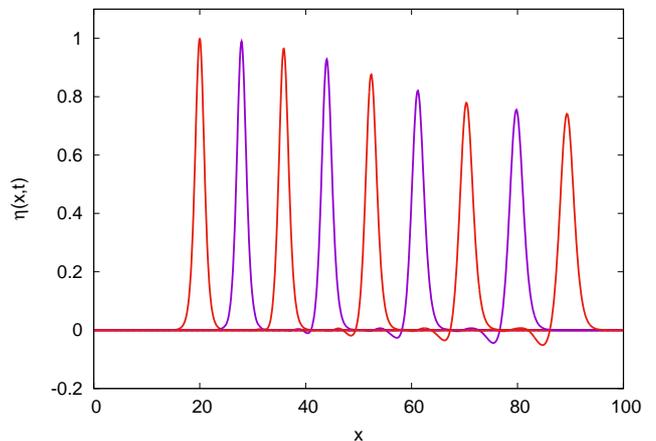}}
\end{center}
\caption{Time evolution of the initial KdV soliton obtained with second order equation derived in \cite{KRR,KRI}. The bottom function $h_-(x)$ and the values of parameters $\alpha,\beta,\delta $ are the same as in the Fig.~3 of \cite{FKMY}.} \label{ab1d5-}
\end{figure}

In the case of the decreasing bottom we repeat the calculations with the bottom function $h_-(x)$. The result of our second order numerical time evolution is displayed in Fig.~\ref{ab1d5-}. In our case the soliton amplitude decreases slower than in \cite[Fig.~3]{FKMY} and the depression behind the main wave is much smaller.

\begin{figure}[tbh]
\begin{center}
\resizebox{0.99\columnwidth}{!}{\includegraphics{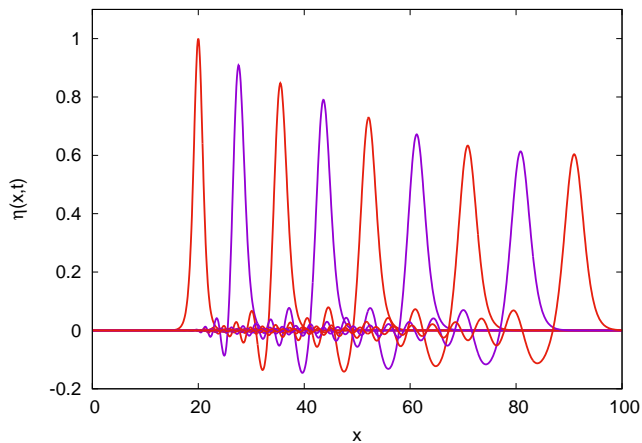}}
\end{center}
\caption{The same as in Fig.~\ref{ab1d5+FKMY} but with the bottom function  $h_-(x)$.} \label{ab1d5-FKMY}
\end{figure}

Next, we repeat the same steps as in the previous subsection. Limiting the equation \cite[Eq.~(40)]{FKMY} to second order, that is, to (\ref{kdv2d}) we obtain with the bottom function $h_-$ the evolution displayed in Fig.~\ref{ab1d5-FKMY} which looks incorrect. 

\begin{figure}[tbh]
\begin{center}
\resizebox{0.99\columnwidth}{!}{\includegraphics{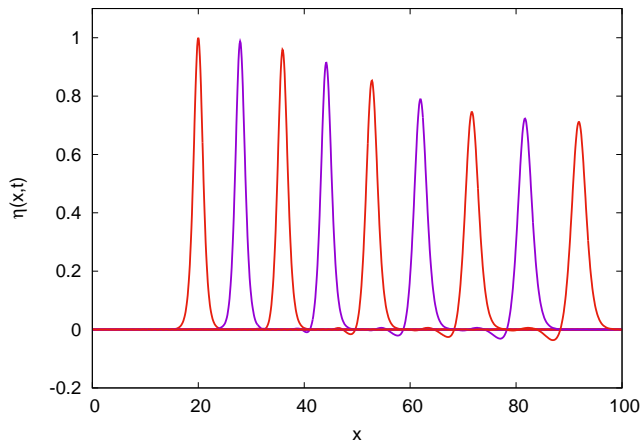}}
\end{center}
\caption{The same as in Fig.~\ref{ab1d5+FKMYcB} but with the bottom function  $h_-(x)$.} \label{ab1d5-FKMYcB}
\end{figure}

Now, as in the case of the ascendant bottom, we replace the incorrect term $\frac{1}{3}\beta \eta_{3x}$ by the correct one,  $\frac{1}{6}\beta \eta_{3x}$. The evolution according to (\ref{kdv2d}) in which the term $\frac{1}{6}\beta \eta_{3x}$ is the right one gives the result presented in Fig.~\ref{ab1d5-FKMYcB}. 
The wave profiles in Figs.~\ref{ab1d5-} and \ref{ab1d5-FKMYcB} differ only very little. This means that, the main reason for the wrong results presented in Figs.~2 and 3 of \cite{FKMY} is the use of the incorrect first order term in the form $\frac{1}{3}\beta \eta_{3x}$. Using the incorrect form of $\beta\delta (\cdots)$ terms has a much smaller influence since these terms are small and the coefficients used in \cite[Eq.~(40)]{FKMY} are only slightly different from the correct ones (compare (\ref{bede}) and (\ref{bedeC})).

\subsection{Surface tension effects}

In order to check the results presented in \cite{FKMY} for nonzero surface tension let us rewrite the equation \cite[Eq.~(42)]{FKMY} but limiting it to second order
\begin{align} \label{kdv2st} 0 &=  \eta_t+\eta_x + \frac{3}{2}\alpha \eta\eta_x +\frac{1}{6}\beta(1-3\tau) \eta_{3x} - \frac{3}{8}\alpha^2\eta^2\eta_x  \hspace{5ex}   \\ &  + \!\alpha\beta\!\left(\!\frac{23\!+\!15\tau}{24}\eta_x\eta_{2x}\!\!+\!\!\frac{5\!-\!3\tau}{12}\eta\eta_{3x}\! \!\right) 
\!+\!\beta^2\!\frac{19\!-\!30\tau\!-\!45\tau^2}{360}\eta_{5x} \nonumber \\ & +\beta\delta \left(\frac{13}{72}h_{3x}\eta + \frac{1}{8}h_{2x}\eta_x -\frac{7}{24}h_{x}\eta_{2x} -\frac{17}{72}h\eta_{3x} \right.  \nonumber  \\ & \left. \hspace{7ex}
-\frac{2}{3\beta} (h_{x}\eta + h\eta_{x})\right). \nonumber 
\end{align}

\begin{figure}[tbh]
\begin{center}
\resizebox{0.99\columnwidth}{!}{\includegraphics{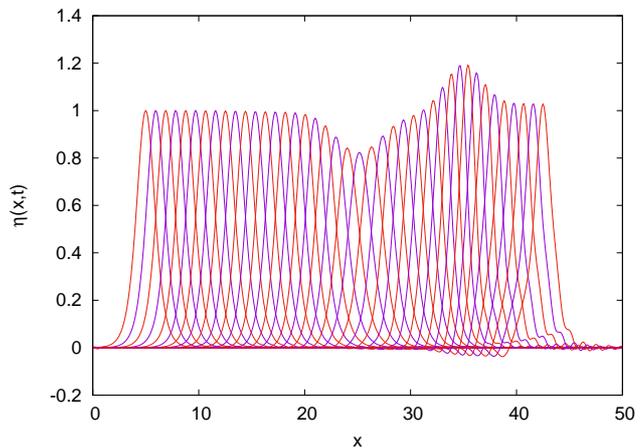}}
\end{center}
\caption{The result of time evolution according to second order equation (\ref{kdv2st}) with $\tau=0$ to compare with \cite[Fig.~6]{FKMY}.} \label{ab1d3-FKMY6}
\end{figure}

In Fig.~\ref{ab1d3-FKMY6} we present the time evolution of the initial KdV soliton according to the equation (\ref{kdv2st}) for all conditions as those used by the authors of \cite{FKMY} in their Fig.~6, that is, for $\alpha=\beta=0.1$, $\delta=0.3$ and the same bottom function.

Comparing this result with \cite[Fig.~6]{FKMY} we see that in our case the changes of the amplitude are much smaller and second order effects are much 'cleaner' than those contained in Fig.~6 of \cite{FKMY}.

\begin{figure}[tbh]
\begin{center}
\resizebox{0.99\columnwidth}{!}{\includegraphics{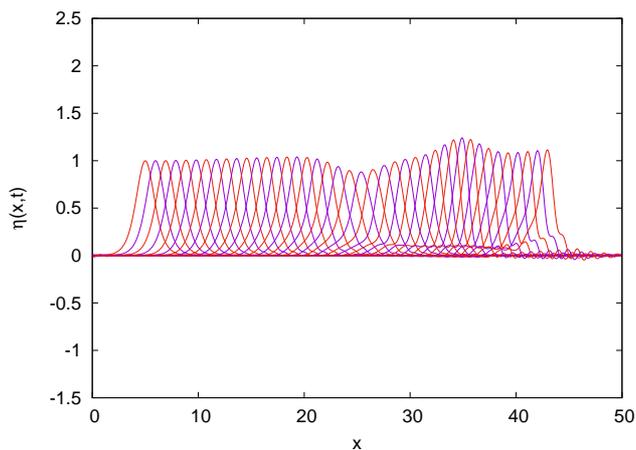}}
\end{center}
\caption{The result of time evolution according to second order equation (\ref{kdv2st}) with $\tau=0.1$ to compare with \cite[Fig.~10]{FKMY} in the same vertical scale.} 
\label{ab1d3-FKMY10}
\end{figure}

Now, we check the influence of the surface tension. In Fig.~10 of  \cite{FKMY}
the authors present the time evolution of the initial KdV soliton according to their third-order equation~(42) in which third order terms are inconsistent. The parameters used were: $\alpha=\beta=0.1$, $\delta=0.3$ and $\tau=0.1$. The bottom was taken as the Gaussian well followed by the symmetric Gaussian hump. 

In Fig.~\ref{ab1d3-FKMY10} we show the time evolution of the same soliton but according to the equation  (42) of  \cite{FKMY} limited to second order, that is, the equation (\ref{kdv2st}). The result is obtained with the same parameters $\alpha,\beta,\delta,\tau$ and the same bottom function. For comparison with the Fig.~10 of \cite{FKMY} we set the same vertical scale.

From Fig.~\ref{ab1d3-FKMY10} it is clear that all rapid oscillations seen in  Fig.~10 of \cite{FKMY} for $30<x<40$ are not present in the evolution according to consistent second order KdV equation~(\ref{kdv2st}). We conclude that the strange behaviour of the soliton motion over bottom changes presented by the authors of  \cite{FKMY} is the result of an inconsistent derivation of third order equation.

\section{Conclusions}

From this short study the following conclusions may be drawn.

\begin{itemize}
\item Consistent incorporation of small amplitude bottom changes into KdV equation is possible exclusively in the perturbation approach of second order with respect to small parameters. This is because the bottom kinetic boundary equation (\ref{botBC3}) can be resolved for $\phi^{(1)}$ only when it is taken in second order.

\item Derivation of the third order KdV equation for the case of uneven bottom 
presented in  \cite{FKMY} is inconsistent since one of equations of the  Euler set of equations is taken in second order and the other ones in third order. 
Therefore these equations are flawed.

\item Numerical tests made by us with consistent second order KdV equation for the case of a non-flat bottom show the absence of any strange behaviour of soliton evolution obtained in \cite{FKMY} with inconsistent third order equations.
\end{itemize}


\begin{thebibliography}{99}

\bibitem{FKMY} M. Fokou, T.C. Kofan\'e, A. Mohamadou and E. Yomba, The third-order perturbed Korteweg-de Vries equation for shallow water waves with a non-flat bottom, Eur. Phys. J. Plus, \textbf{132}, 410 (2017).

\bibitem{MS90} T.R. Marchant and N.F. Smyth, The extended Korteweg--de Vries equation and the resonant flow of a fluid over topography,  J. Fluid Mech.  \textbf{221}, 263-288 (1990).

\bibitem{KRR} A. Karczewska, P. Rozmej and  Ł. Rutkowski,
A new nonlinear equation in the shallow water wave problem,
Physica Scripta, \textbf{89}, 054026 (2014).

\bibitem{KRI} A. Karczewska, P. Rozmej and E. Infeld, Shallow-water soliton dynamics beyond the Korteweg - de Vries equation, Phys. Rev. E \textbf{90}, 012907 (2014).

\bibitem{BS13} G.I. Burde and A. Sergyeyev,
Ordering of two small parameters in the shallow water wave problem,
J. Phys. A: Math. Theor. \textbf{46}, 075501 (2013). 

\bibitem{FKMY16} M. Fokou, T.C. Kofan\'e, A. Mohamadou and E. Yomba, One- and two-soliton solutions to a new KdV evolution equation with nonlinear and nonlocal terms for the water wave problem,
Nonlinear Dyn., \textbf{83} (4), 2461-2473 (2016).


\bibitem{Daw} S.P. Dawson, Solitons and radiation described by the derivative nonlinear Schr\"odinger equation., Phys. Rev. A, \textbf{45}, 7448-7455 (1995). 

\bibitem{Kuz} E.A. Kuznetsov,  A.V. Mikhailov, I.A. Shimokhin, Nonlinear interaction of solitons and radiation, Physica D, \textbf{87}, 201-215 (1995).

\bibitem{GrimPoF}  R.H.J. Grimshaw, K.H. Chan and K.W. Chow, Transcritical flow of a stratified fluid: The forced extended Korteweg–de Vries model, Physics of Fluids, \textbf{14} (2), 755-774 (2002).
\end{thebibliography}
\end{document}